\documentclass[preprint,aps,prd,showpacs,groupedaddress,superscriptaddress,%
floatfix,nofootinbib]{revtex4}%
\usepackage{graphicx}
\usepackage{amsmath}
\usepackage{bm}
\begin{document}
\preprint{ANL-HEP-PR-07-48}
\title{Improved determination of color-singlet \\
nonrelativistic QCD matrix elements for $\bm{S}$-wave charmonium
}
\author{Geoffrey~T.~Bodwin}
\affiliation{High Energy Physics Division, Argonne National Laboratory,\\
9700 South Cass Avenue, Argonne, Illinois 60439, USA}
\author{Hee~Sok~Chung}
\affiliation{Department of Physics, Korea University, Seoul 136-701, Korea}
\author{Daekyoung~Kang}
\affiliation{Department of Physics, Korea University, Seoul 136-701, Korea}
\affiliation{Physics Department, Ohio State University, 
Columbus, Ohio 43210, USA}
\author{Jungil~Lee}
\affiliation{High Energy Physics Division, Argonne National Laboratory,\\
9700 South Cass Avenue, Argonne, Illinois 60439, USA}
\affiliation{Department of Physics, Korea University, Seoul 136-701, Korea}
\author{Chaehyun~Yu}
\affiliation{Department of Physics, Korea University, Seoul 136-701, Korea}

\begin{abstract}
We present a new computation of $S$-wave color-singlet nonrelativistic
QCD matrix elements for the $J/\psi$ and the $\eta_c$. We compute the
matrix elements of leading order in the heavy-quark velocity $v$ and the
matrix elements of relative order $v^2$. Our computation is based on the
electromagnetic decay rates of the $J/\psi$ and the $\eta_c$ and on a
potential model that employs the Cornell potential. We include
relativistic corrections to the electromagnetic decay rates, resumming a
class of corrections to all orders in $v$, and find that they
significantly increase the values of the matrix elements of leading
order in $v$. This increase could have important implications for
theoretical predictions for a number of quarkonium decay and production
processes. The values that we find for the matrix elements of relative
order $v^2$ are somewhat smaller than the values that one obtains from
estimates that are based on the velocity-scaling rules of
nonrelativistic QCD.
\end{abstract}

\maketitle
\section{Introduction\label{sec:intro}}
In the nonrelativistic quantum chromodynamics (NRQCD)
factorization formalism  \cite{Bodwin:1994jh},
heavy-quarkonium decay and production rates are expressed as
sums of short-distance coefficients times 
NRQCD operator matrix elements. The matrix elements in
these sums scale as powers of $v$, the typical heavy-quark (or
antiquark) velocity in the quarkonium rest frame. Hence, the sum in the
NRQCD factorization expression can be thought of as an expansion in
powers of $v$. The term that is proportional to the matrix
element of leading order in $v$ often gives the dominant contribution in
decay and production processes. The leading-order matrix element
involves the production or annihilation of a heavy quark-antiquark 
($Q\bar Q$) pair in a color-singlet state. The term that is
proportional to the matrix element of relative order $v^2$ gives the
first relativistic correction. This order-$v^2$ matrix element also
involves the production or annihilation of a heavy $Q\bar Q$ 
pair in a color-singlet state.

In the vacuum-saturation approximation \cite{Bodwin:1994jh} for decay
matrix elements, one keeps only the vacuum intermediate state, while, in
the vacuum-saturation approximation for production matrix elements, one
keeps only the heavy $Q\bar Q$ intermediate state. The
vacuum-saturation approximation is valid up to corrections of relative
order $v^4$ \cite{Bodwin:1994jh}. In this approximation, the
color-singlet decay matrix elements are equal to color-singlet
production matrix elements. These vacuum-saturation matrix elements are
also the relevant ones for purely electromagnetic decay and production
processes and for exclusive decay and production processes. The
vacuum-saturation matrix element at leading order in $v$ is proportional
to the square of the quarkonium wave function at the origin. In this
paper, we compute the vacuum-saturation matrix elements of leading order
in $v$ and of relative order $v^2$ for the $J/\psi$ and $\eta_c$ states.

The analysis of these matrix elements for the $J/\psi$ state differs in
several respects from a previous one involving some of the authors
\cite{Bodwin:2006dn}. In that analysis, the matrix element at leading
order in $v$ was obtained by comparing the theoretical expression for
the decay width $\Gamma[J/\psi \to e^+e^-]$ with the experimental
measurement. The theoretical expression that was used in that
analysis included the order-$\alpha_s$ correction, but not the
relativistic corrections. In the present paper, we include those
relativistic corrections. The matrix element of relative order $v^2$ is
determined from a potential-model calculation \cite{Bodwin:2006dn} that
uses the leading-order matrix element as an input. The relativistic
corrections to $\Gamma[J/\psi \to e^+e^-]$ in turn depend upon that
order-$v^2$ matrix element. 
Hence, the leading-order and order-$v^2$ matrix elements
are related through a coupled pair of nonlinear equations, which we
solve numerically.

We obtain values for the $\eta_c$ matrix elements in two different
ways and average the results. First, we obtain a set of values by making
use of the comparison between theory and experiment for the width
$\Gamma[\eta_c\to\gamma\gamma]$. This comparison gives one nonlinear
equation for the matrix elements. As in the $J/\psi$ case, we make use
of a potential-model calculation of the order-$v^2$ matrix element to
obtain a second nonlinear equation, and we solve the coupled nonlinear
equations numerically to obtain a set of values for the $\eta_c$ matrix
elements. We obtain a second set of values by making use of the fact
that, because of the approximate heavy-quark spin symmetry of NRQCD
\cite{Bodwin:1994jh}, the $\eta_c$ and $J/\psi$ matrix elements are
equal, up to corrections of relative order $v^2$. We define this second
set of values for the $\eta_c$ matrix elements simply by taking the
values that we obtain for the $J/\psi$ matrix elements and appending
additional error bars that take into account the order-$v^2$ corrections
to the heavy-quark-spin-symmetry relation. 

Because the two sets of values for the $\eta_c$ matrix elements that we
obtain in this way have input parameters (such as the heavy-quark mass
and the string tension) in common, the uncertainties in these matrix
elements are highly correlated between sets and between matrix elements
within a set. Therefore, we carry out a covariance-matrix analysis to
compute the average. The $J/\psi$ and $\eta_c$ matrix
elements are also highly correlated. Such correlations could be
important in applications of our results to calculations involving both
the $J/\psi$ and the $\eta_c$ and/or order-$v^2$ corrections.
Therefore, we present tables showing the variations of each of the
matrix elements with respect to the various sources of uncertainty
and also give the covariance matrix that corresponds to these variations.

A further refinement that we include in this work is to resum a class of
relativistic corrections \cite{Bodwin:2006dn} to $\Gamma[J/\psi \to
e^+e^-]$ and $\Gamma[\eta_c\to\gamma\gamma]$. First, we consider all
corrections that arise from matrix elements involving only color-singlet
$Q\bar Q$ Fock states. By making use of a generalization of the
Gremm-Kapustin relation \cite{Gremm:1997dq,Bodwin:2006dn}, we can
determine all of these matrix elements, up to corrections of relative
order $v^2$, from the leading-order and order-$v^2$ matrix elements. The
simple expressions that result can easily be summed to all orders in
$v$. This resummation is equivalent to retaining all of the relativistic
corrections that are contained in a potential-model $Q\bar Q$ wave
function, up to the ultraviolet cutoff of the NRQCD matrix elements.

Because the expressions for the matrix elements of order $v^2$ and
higher are accurate only up to corrections of relative order $v^2$, the
uncertainty in the resummed expression is of order $v^4$
relative to the leading-order expression. That is, the nominal accuracy
in $v$ is no higher than that of a fixed-order calculation through
relative order $v^2$. However, if the relativistic corrections to a
given process that arise from the $Q\bar Q$ Fock-state wave function have
particularly large coefficients in the $v$ expansion, then the use of
the resummed expression may improve the numerical accuracy. Furthermore,
the resummation may give an indication of the rate of convergence of
the $v$ expansion. In any case, it is generally desirable to include in a
calculation a well-defined, if incomplete, set of contributions whenever
possible.

The remainder of this paper is organized as follows.  In
Sec.~\ref{sec:me}, we review the definitions of the $S$-wave NRQCD
matrix elements at the leading and higher orders in $v$, and we give
the relations of these matrix elements to the quarkonium wave functions.
We also introduce the generalized Gremm-Kapustin relation for an $S$-wave
quarkonium state, which expresses  matrix elements of higher order in $v$
in terms of the matrix element of leading order in $v$ and the binding
energy. The generalized Gremm-Kapustin relation allows us to resum a
class of relativistic corrections to quarkonium decay to all orders in
$v$. In Sec.~\ref{sec:resum}, we present the resummed formulas for the
electromagnetic decay widths of the $J/\psi$ and the $\eta_c$.
Sec.~\ref{sec:cornell} contains a description of the potential-model
method that we use to compute the binding energy of the $S$-wave states
and, through the generalized Gremm-Kapustin relation, the NRQCD matrix
elements of higher order in $v$. In Sec.~\ref{sec:det}, we compute the
numerical values of the NRQCD matrix elements for the $J/\psi$ and the
$\eta_c$. We compare our results for the matrix elements with those from
previous determinations in Sec.~\ref{sec:comparisons}. Finally, we
summarize our results in Sec.~\ref{sec:conclusions}.

\section{NRQCD matrix elements\label{sec:me}}
\subsection{Decay and production matrix elements}

In the cases of the inclusive decays of spin-singlet and spin-triplet $S$-wave
quarkonium states, such as the $\eta_c$ and the $J/\psi$, the matrix
elements at the leading power in $v$ are
\begin{subequations}
\label{decay-me}%
\begin{eqnarray}
\langle\mathcal{O}_1({}^1S_0)\rangle_{H}&=&
\langle 
H({}^1S_0)|\psi^\dagger\chi\chi^\dagger\psi|H({}^1S_0)\rangle,\\
\langle\mathcal{O}_1({}^3S_1)\rangle_{H}&=&
\langle H({}^3S_1)|\psi^\dagger\bm{\sigma}\chi\cdot
\chi^\dagger\bm{\sigma}\psi|H({}^3S_1)\rangle,
\end{eqnarray}
\end{subequations}
where $H$ is a quarkonium state, ${}^{2s+1}S_J$ is the standard
spectroscopic notation for a state with spin angular momentum $s$,
 orbital angular
momentum zero, and total angular momentum $J$, $\psi$ is a two-component
Pauli spinor that annihilates a heavy quark, $\chi$ is a
two-component Pauli spinor that creates a heavy antiquark, and $\sigma^i$
is a Pauli matrix. The subscript $1$ on an NRQCD operator $\mathcal{O}$
indicates that it is a color-singlet operator. 

Similarly, in the case of
the inclusive production of spin-singlet and spin-triplet $S$-wave
quarkonium states, the matrix elements at the leading power in $v$ are
\begin{subequations}
\label{prod-me}%
\begin{eqnarray}
\langle\mathcal{O}_1({}^1S_0)_H\rangle&=&
\langle 0|\chi^\dagger\psi
\left(\sum_{X,\;\hbox{pol.}} |H({}^1S_0)+X\rangle\langle H({}^1S_0)+X|\right) 
\psi^\dagger\chi|0\rangle,\\
\langle\mathcal{O}_1({}^3S_1)_H\rangle&=&
\langle 0|\chi^\dagger\sigma^i\psi
\left(\sum_{X,\;\hbox{pol.}} |H({}^3S_1)+X\rangle\langle H({}^3S_1)+X|\right) 
\psi^\dagger\sigma^i\chi|0\rangle,
\end{eqnarray}
\end{subequations}
where the sum is over the light degrees of freedom $X$
and the $2J+1$ quarkonium polarizations.

In the vacuum-saturation approximation \cite{Bodwin:1994jh}, which is
valid up to corrections of relative order $v^4$, the decay matrix
elements in Eq.~(\ref{decay-me}) and $1/(2J+1)$ times the production
matrix elements in Eq.~(\ref{prod-me}) both reduce to
\begin{subequations}                                                  
\label{VS-me}%
\begin{eqnarray}
\langle\mathcal{O}_1({}^1S_0)\rangle_{H}^{\textrm{VS}}&=&
|\langle 0|\chi^\dagger\psi|H({}^1S_0)\rangle|^2,\\
\langle\mathcal{O}_1({}^3S_1)\rangle_{H}^{\textrm{VS}}&=&
|\langle 0|\chi^\dagger\bm{\sigma}\psi|H({}^3S_1)\rangle|^2\nonumber\\
&=&|\langle 0|\chi^\dagger\bm{\sigma}\cdot\bm{\epsilon}^*
\psi|H({}^3S_1)\rangle|^2,
\label{VS-me3}%
\end{eqnarray}                                                      
\end{subequations}
in the spin-singlet and spin-triplet cases, respectively. In
Eq.~(\ref{VS-me3}), the quarkonium polarization vector is denoted by
$\epsilon$, and there is no sum over the polarization states of the
quarkonium.  In the cases of purely electromagnetic decay or production
or exclusive decay or production, the matrix elements in
Eq.~(\ref{VS-me}) are the relevant ones at leading order in $v$.

The first relativistic corrections to inclusive $S$-wave decay and
production involve operators that are analogous to those in
Eqs.~(\ref{decay-me}) and (\ref{prod-me}), but that contain a factor
of $(-\tfrac{i}{2}\tensor{\bm{D}})^2$ between either $\psi^\dagger$
and $\chi$ or $\chi^\dagger$ and $\psi$. Here, $\tensor{\bm{D}}$ is
the spatial part of the covariant derivative acting to the left and
right anti-symmetrically: $\chi^\dagger \tensor{\bm{D}}\psi \equiv
\chi^\dagger ({\bm{D}} \psi) - ({\bm{D}} \chi)^\dagger \psi$. These
operators are of order $v^2$ relative to those in Eqs.~(\ref{decay-me})
and (\ref{prod-me}). The corresponding matrix elements 
reduce in the vacuum-saturation approximation to
\begin{subequations}                                                  
\label{VS-me-vsq}%
\begin{eqnarray}
\langle\mathcal{P}_1({}^1S_0)\rangle_{H}^{\textrm{VS}}&=&
\textrm{Re\,} \big[
\langle H({}^1S_0)|\psi^\dagger\chi|0\rangle
\langle 0|
\chi^\dagger(-\tfrac{i}{2}\tensor{\bm{D}})^2\psi
|H({}^1S_0)\rangle
\big],\\
\langle\mathcal{P}_1({}^3S_1)\rangle_{H}^{\textrm{VS}}&=&
\textrm{Re\,} \big[
\langle H({}^3S_1)|\psi^\dagger\bm{\sigma}\cdot\bm{\epsilon}\chi|0\rangle
\langle 0|\chi^\dagger\bm{\sigma}\cdot\bm{\epsilon}^*
(-\tfrac{i}{2}\tensor{\bm{D}})^2\psi|H({}^3S_1)\rangle
\big].
\end{eqnarray}
\end{subequations}
In the cases of purely electromagnetic production or decay or exclusive 
production or decay, the matrix elements in Eq.~(\ref{VS-me-vsq}) are 
the relevant ones. 

Corrections of still higher orders in $v^2$ involve, among other matrix 
elements, those in which higher powers of $(-\tfrac{i}{2}\tensor{\bm{D}})^2$ 
appear. It is convenient to define ratios of these 
matrix elements to the matrix elements of lowest order in $v$ :
\begin{subequations}
\label{me-ratios}%
\begin{eqnarray}
\langle \bm{q}^{2r}\rangle_{H({}^1S_0)}&=&
\frac{\langle 0|\chi^\dagger(-\tfrac{i}{2}\tensor{\bm{D}})^{2r}\psi
|H({}^1S_0)\rangle}
{\langle 0|\chi^\dagger\psi|H({}^1S_0)\rangle},\\
\langle \bm{q}^{2r}\rangle_{H({}^3S_1)}&=&
\frac{\langle 0|\chi^\dagger\bm{\sigma}\cdot\bm{\epsilon}^*
(-\tfrac{i}{2}\tensor{\bm{D}})^{2r}\psi|H({}^3S_1)\rangle}
{\langle 0|
\chi^\dagger\bm{\sigma}\cdot\bm{\epsilon}^*\psi
|H({}^3S_1)\rangle},
\end{eqnarray}
\end{subequations}
where $\bm{q}$ is half the relative three-momentum of 
the $Q$ and $\bar{Q}$ in the quarkonium rest frame.

In this paper, we compute the quantities
$\langle\mathcal{O}_1({}^1S_0)\rangle^{\textrm{VS}}_{\eta_c}$ and
$\langle\mathcal{O}_1({}^3S_1)\rangle^{\textrm{VS}}_{J/\psi}$, which
are given by Eq.~(\ref{VS-me}), and the quantities $\langle
\bm{q}^{2r}\rangle_{\eta_c}$ and $\langle
\bm{q}^{2r}\rangle_{J/\psi}$, which are given by
Eq.~(\ref{me-ratios}). As we shall see, the higher-order ratios in
Eq.~(\ref{me-ratios}) can be related to the lowest-order ones by making
use of a generalization of the Gremm-Kapustin relation
\cite{Gremm:1997dq,Bodwin:2006dn}.

As is discussed in Ref.~\cite{Bodwin:2006dn}, the higher-order matrix
elements in Eq.~(\ref{me-ratios}) contain power ultraviolet divergences
and require regularization. In this paper, we regulate these power
divergences dimensionally at the one-loop level. One-loop dimensional
regularization of the matrix elements is appropriate for use in
conjunction with one-loop calculations of the
short-distance coefficients.

\subsection{Relations of NRQCD matrix elements to quarkonium wave functions
\label{sec:relations-me-wf}}

In the rest frame of an $S$-wave heavy quarkonium $H$ in a spin-singlet 
($^1S_0$) or spin-triplet ($^3S_1$) 
state, one can express the wave function at the origin of the leading
$Q\bar{Q}$ Fock state in terms of the color-singlet
NRQCD matrix elements~\cite{Bodwin:1994jh}:
\begin{subequations}
\label{psi0}%
\begin{eqnarray}
\psi_{H({}^1S_0)}(0)&=&
\int\frac{d^3q}{(2\pi)^3}
\widetilde{\psi}_{H({}^1S_0)}(\bm{q})
=\frac{1}{\sqrt{2N_c}}
\langle 0|\chi^\dagger\psi|H({}^1S_0)\rangle,
\\
\bm{\epsilon}\psi_{H({}^3S_1)}(0)&=&\bm{\epsilon}\int\frac{d^3q}{(2\pi)^3}
\widetilde{\psi}_{H({}^3S_1)}(\bm{q})
=\frac{1}{\sqrt{2N_c}}
\langle 0|\chi^\dagger\bm{\sigma}\psi|H({}^3S_1)\rangle.
\end{eqnarray}
\end{subequations}
$\widetilde{\psi}_{H}(\bm{q})$ is the momentum-space wave function
for the leading $Q(\bm{q})\bar{Q}(-\bm{q})$ Fock state of the 
quarkonium.  The wave function is, of course, gauge dependent.
Throughout this paper, we work in the Coulomb gauge. The 
normalization factor $1/\sqrt{2N_c}$ accounts for the traces in the
SU(2)-spin and SU(3)-color spaces. 

Relativistic corrections to the production and decay rates for
a heavy quarkonium involve matrix elements that are related to 
derivatives of the wave function at the origin:
\begin{subequations}
\begin{eqnarray}
\psi^{(2r)}_{H({}^1S_0)}(0)&\equiv&
\int\frac{d^3q}{(2\pi)^3}
\bm{q}^{2r}
\widetilde{\psi}_{H({}^1S_0)}(\bm{q})
=\frac{1}{\sqrt{2N_c}}
\langle 0|\chi^\dagger
(-\tfrac{i}{2}\tensor{\bm{\nabla}})^{2r}
\psi|H({}^1S_0)\rangle,
\\
\bm{\epsilon}\psi_{H({}^3S_1)}^{(2r)}(0)
&\equiv&\bm{\epsilon}\int\frac{d^3q}{(2\pi)^3}
\bm{q}^{2r}
\widetilde{\psi}_{H({}^3S_1)}(\bm{q})
=\frac{1}{\sqrt{2N_c}}
\langle 0|\chi^\dagger\bm{\sigma}
(-\tfrac{i}{2}\tensor{\bm{\nabla}})^{2r}
\psi|H({}^3S_1)\rangle.
\end{eqnarray}
\label{psi20}%
\end{subequations}
Usually, these operator matrix elements are written in terms of the
covariant derivative $\tensor{\bm{D}}$ (Ref.~\cite{Bodwin:1994jh}), as
in Eqs.~(\ref{VS-me-vsq}) and (\ref{me-ratios}), rather than
$\tensor{\bm{\nabla}}$. However, in the Coulomb gauge, the difference
between the $\tensor{\bm{D}}$ and $\tensor{\bm{\nabla}}$ is suppressed
as $v^2$ (Ref.~\cite{Bodwin:1994jh}). We emphasize again that the
derivatives of the wave function at the origin, as defined in
Eq.~(\ref{psi20}), are ultraviolet-divergent quantities, which must
be regulated. We also note that $\psi_{H}^{(2r)}(0)$ is different from the
expectation value of $\bm{q}^{2r}$:
\begin{equation}
\psi^{(2r)}_{H}(0)\neq
\int\frac{d^3q}{(2\pi)^3}
\,\bm{q}^{2r}
\widetilde{\psi}_{H}^*(\bm{q})
\widetilde{\psi}_{H}(\bm{q}).
\end{equation}

Comparing Eq.~(\ref{psi20}) with Eq.~(\ref{me-ratios}), we see that
\begin{equation}
\langle \bm{q}^{2r}\rangle_H=\frac{\psi^{(2r)}_{H}(0)}{\psi_{H}(0)}
[1+\mathcal{O}(v^2)].
\label{p-sq}%
\end{equation}
One can also define matrix elements of powers of the heavy-quark
velocity in terms of matrix elements of powers of the heavy-quark
momentum:
\begin{equation}
\langle v^{2r}\rangle_{H}=
\langle \bm{q}^{2r}\rangle_{H}/m_Q^{2r},
\label{v-sq}%
\end{equation}
where $m_Q$ is the heavy-quark mass.

\subsection{The generalized Gremm-Kapustin relation%
\label{sec:direct}}
From the effective field theory known as potential NRQCD (pNRQCD)
\cite{Brambilla:1999xf}, it follows that one can compute the wave
functions at the origin and derivatives of wave functions at the origin
in Eq.~(\ref{p-sq}), up to errors of relative order $v^2$, from the
Schr\"odinger wave function for a heavy $Q\bar Q$ pair interacting
through the leading (static) spin-independent $Q\bar Q$ potential.
It was shown in Ref.~\cite{Bodwin:2006dn} that, in the case of a
spin-independent potential and for dimensionally regulated matrix
elements, the ratios in Eq.~(\ref{p-sq}) are related through the
generalized Gremm-Kapustin relation:
\begin{equation}
\langle \bm{q}^{2r} \rangle_{H}
=(m\,\epsilon_{nS})^{r}
[1+\mathcal{O}(v^2)],
\label{gremm-kapustin}%
\end{equation}
where $\epsilon_{nS}$ is the binding energy of the $Q\bar Q$ pair in the
quarkonium state $H$ with principal quantum number $n$ and orbital
angular momentum $S$, and $m$ is the heavy-quark mass in the effective
theory pNRQCD. The relation (\ref{gremm-kapustin}) follows from the
equations of motion of the $Q\bar Q$ pair and from dimensional
regularization of the matrix elements at the one-loop level, provided
that the potential is parametrized as a sum of constants times powers of
the $Q\bar Q$ separation. The Cornell potential \cite{Eichten:1978tg},
which we will employ later in this paper, is parametrized in this way.
We note that Eq.~(\ref{gremm-kapustin}) implies that
\begin{equation}
\langle \bm{q}^{2r}\rangle_H=\langle \bm{q}^2\rangle_H^r,
\label{g-k-2}%
\end{equation}
up to corrections of relative order $v^2$.

We will use Eq.~(\ref{gremm-kapustin}) to determine the quantities
$\langle \bm{q}^{2r}\rangle_{\eta_c}$ and $\langle
\bm{q}^{2r}\rangle_{J/\psi}$. 
In order to evaluate the ground-state binding energy
$\epsilon_{1S}$, we will make use of a potential model that is based on
the Cornell potential.

\section{Formulas for electromagnetic decays of
$\bm{S}$-wave heavy quarkonia\label{sec:resum}}
In this section, we present the NRQCD factorization expressions for the 
electromagnetic decay widths $\Gamma[H({}^3S_1)\to e^+e^-]$ and 
$\Gamma[H({}^1S_0)\to\gamma\gamma]$. In subsequent parts of this paper, 
we will apply these formulas to the decays $J/\psi\to e^+e^-$ and 
$\eta_c\to \gamma\gamma$.

\subsection{$\bm{\Gamma[H({}^3S_1)\to e^+e^-]}$\label{sec:resum-psi}}
The NRQCD factorization formula for the amplitude for the decay 
$H({}^3S_1)\to e^+e^-$ is
\begin{equation}
\mathcal{A}[H({}^3S_1)\to e^+ e^-]
=
\sqrt{2m_H}
\sum_n d_n({}^3S_1) \langle 0|\mathcal{O}_n|H({}^3S_1)\rangle,
\label{A3H-full}%
\end{equation}
where $m_H$ is the quarkonium mass, the $d_n({}^3S_1)$ are
short-distance coefficients, and the $\mathcal{O}_n$ are NRQCD
operators. The prefactor $\sqrt{2m_H}$ compensates for the fact that the
hadronic NRQCD operator matrix elements conventionally have
nonrelativistic normalization, while we choose the amplitude on the left
side of Eq.~(\ref{A3H-full}) to have relativistic normalization.

Now we approximate the formula (\ref{A3H-full}) by retaining only 
those operator matrix elements that connect the vacuum to the 
color-singlet, $Q\bar Q$ Fock state of the quarkonium $H$. Then, we 
have
\begin{equation}
\mathcal{A}[H({}^3S_1)\to e^+e^-]
=
\sqrt{2m_H}
\sum_n c^i_n({}^3S_1) \langle 0|
\chi^\dagger (-\tfrac{i}{2}\tensor{\bm{D}})^{2n} \sigma^i
\psi|H({}^3S_1)\rangle,
\label{A3H}%
\end{equation}
where the short-distance coefficients $c^i_n({}^3S_1)$ are a subset of 
the short-distance coefficients $d_n$ in Eq.~(\ref{A3H-full}). We will 
clarify the meaning of the approximation that we have taken to arrive at 
Eq.~(\ref{A3H}) below.

Because the $c^i_n({}^3S_1)$ are insensitive to the long-distance nature
of the hadronic state, we can calculate them by replacing the initial
hadronic state $\sqrt{2m_H}|H({}^3S_1)\rangle$ in Eq.~(\ref{A3H}) with a
perturbative spin-triplet $S$-wave $Q\bar{Q}$ state:
\begin{equation}
\mathcal{A}[Q\bar{Q}_1({}^3S_1)\to e^+e^-]
=
\sum_n
c^i_n({}^3S_1)
\langle 0|\chi^\dagger 
(-\tfrac{i}{2}\tensor{\bm{D}})^{2n} \sigma^i
\psi|Q\bar{Q}_1({}^3S_1)\rangle.
\label{A3Q}%
\end{equation}
The factor $\sqrt{2m_H}$ is absent in Eq.~(\ref{A3Q}) because we
use the same (relativistic) normalization for the $Q\bar{Q}$ state
on both sides of Eq.~(\ref{A3Q}).

In the rest frame of the quarkonium, the perturbative amplitude on the
left side of Eq.~(\ref{A3Q}) at order $\alpha_s^{0}$ is
\cite{Bodwin:2002hg}
\begin{equation}
\mathcal{A}[Q\bar{Q}_1({}^3S_1)\to e^+e^-]
=
\sqrt{2N_c}\, 2E(q)\frac{e^2e_Q}{m_H^2}\,
\bm{L}\cdot\bm{\epsilon}
\bigg( 1 -\frac{\bm{q}^2}{3E(q)[E(q)+m_Q]} \bigg),
\label{A3Q-F}%
\end{equation}
where $e$ is the electromagnetic coupling constant, $e_Q$ is the
electric charge of the heavy quark, $\epsilon$ is the polarization vector for
the spin-triplet state, $L$ is the leptonic current, and
$E(q)=\sqrt{m_Q^2+\bm{q}^2}$. In the expression (\ref{A3Q-F}), we have
neglected the electron mass in comparison with the quarkonium mass. The
factor $1/m_H^2$ arises from the photon propagator. The perturbative
matrix elements on the right side of Eq.~(\ref{A3Q}) are given by
\begin{equation}
\langle 0|\chi^\dagger 
(-\tfrac{i}{2}\tensor{\bm{D}})^{2n}\sigma^i
\psi|Q\bar{Q}_1({}^3S_1)\rangle
=
\sqrt{2N_c}\, 2E(q)\,\bm{q}^{2n} \epsilon^i.
\label{Qnorm}%
\end{equation}
The factor $2E(q)$ arises from the relativistic normalization of the
$Q\bar{Q}$ state. By comparing Eqs.~(\ref{A3Q}) and (\ref{A3Q-F}), one
can read off the short-distance coefficients $c_n^i({}^3S_1)$:
\begin{equation}
c_n^i({}^3S_1) = 
\frac{e^2e_Q}{m_H^2}\,L^i
\left[
\frac{1}{n!}\left(\frac{\partial }{\partial \bm{q}^2}
\right)^n
\bigg( 1- \frac{\bm{q}^2}{3 E(q) [E(q)+m_Q]} 
\bigg)
\right]_{\bm{q}^2=0}.
\label{c3n}%
\end{equation}

Substituting the $c_n^i({}^3S_1)$ in Eq.~(\ref{c3n}) 
into Eq.~(\ref{A3H}), using
Eq.~(\ref{g-k-2}), and including the order-$\alpha_s$ correction to the
amplitude \cite{Barbieri:1975ki,Celmaster:1978yz}, we obtain
\begin{equation}
\mathcal{A}[H({}^3S_1)\to e^+ e^-]
=
\sqrt{2m_H} 
 \frac{e^2e_Q}{m_H^2}L^i
\left[1-f(\langle \bm{q}^2\rangle_H/m_Q^2)
                   -2C_F\frac{\alpha_s}{\pi}\right]
\langle 0|
\chi^{\dagger}
\sigma^i \psi|H({}^3S_1)\rangle,
\label{A3H-F}%
\end{equation}
where $f(x)$ is defined by
\begin{equation}
f(x)=\frac{x}{3(1+x+\sqrt{1+x})}.
\label{fx}%
\end{equation}

Now we can clarify the meaning of the approximation that was taken to
arrive at Eq.~(\ref{A3H}). Suppose that we specialize to the Coulomb
gauge. Then, we can drop the gauge fields in covariant derivatives in the
matrix elements in Eq.~(\ref{A3H}), making errors of relative order
$v^2$. The matrix elements are then proportional to derivatives of the
Coulomb-gauge color-singlet $Q\bar Q$ quarkonium wave function at the
origin \cite{Bodwin:1994jh}. (See Sec.~\ref{sec:relations-me-wf}.) That is,
they are proportional to the moments of the momentum-space wave function
with respect to the wave-function momentum (the relative momentum of the
$Q$ and $\bar Q$). From Eq.~(\ref{c3n}), we see that the short-distance
coefficients $c_n^i({}^3S_1)$, when contracted into $\epsilon^i$, are
the coefficients of the Taylor expansion of
$\mathcal{A}[Q\bar{Q}_1({}^3S_1)\to e^+e^-]/[\sqrt{2N_c}\,2E(q)]$ with
respect to the wave-function momentum. Hence, Eq.~(\ref{A3H}) has the
interpretation of the convolution of the short-distance amplitude with
the momentum-space quarkonium wave function, where the short-distance
coefficients have been Taylor expanded with respect to the wave-function
momenta. Therefore, we see that the approximate NRQCD expansion in
Eq.~(\ref{A3H}) includes all of the relativistic corrections that are
contained in the color-singlet $Q\bar Q$ quarkonium wave function, up to
the ultraviolet cutoff of the NRQCD matrix elements.\footnote{We note
that, in the case of dimensionally regulated NRQCD matrix elements, pure
power ultraviolet divergences in the matrix elements are set to zero.
Hence, the effects of purely power-divergent contributions are absent in
the resummation.}

We note that, in the quarkonium rest frame, the square of the spatial
part of the leptonic factor $L$, summed over lepton spins, is given by
\begin{equation}
\sum_{\textrm{spins}}L^iL^{*j}
=
2m_H^2\left( \delta^{ij} 
          -\hat{\bm{k}}^i \hat{\bm{k}}^j \right),
\label{L-sq}%
\end{equation}
where $\hat{\bm{k}}=\bm{k}/|\bm{k}|$ and $\bm{k}$ is the
three-momentum of the $e^-$ in the quarkonium rest frame. The temporal 
parts of $\sum_{\textrm{spins}}L^\mu L^{*\nu}$ vanish in the 
quarkonium rest frame.

We obtain the leptonic decay width of the spin-triplet $S$-wave heavy
quarkonium by taking the square of the amplitude (\ref{A3H-F}),
summing over lepton spins using Eq.~(\ref{L-sq}), averaging over
the $H({}^3S_1)$ polarization states, and multiplying by the two-body
phase space and the normalization $(2m_H)^{-1}$. 
The result is \cite{Bodwin:2002hg,Chung:2007ke}
\begin{equation}
\Gamma[H({}^3S_1)\to e^+e^-]
=
\frac{8\pi e_Q^2\alpha^2}{3m_H^2}
\,\left[1-f(\langle \bm{q}^2\rangle_H/m_Q^2)
         -2C_F\frac{\alpha_s}{\pi}\right]^2
\langle \mathcal{O}_1\rangle_H,
\label{G3H-F}%
\end{equation}
where $\alpha=e^2/(4\pi)$.
In Eq.~(\ref{G3H-F}), the explicit relativistic corrections are
contained in the term $-f(\langle \bm{q}^2/m_Q^2\rangle_H)$. In
addition, there are implicit relativistic corrections that are contained
in the factors $m_H$. Strictly speaking, if one were to compute the
decay amplitude completely with the framework of NRQCD, then $m_H$ would
be written as $2E(q)$ and expanded in powers of $|\bm{q}|/m_Q$ to obtain
additional relativistic corrections. (See, for example,
Refs.~\cite{Bodwin:2002hg,Braaten:2002fi}.) However, we note that the
factor $1/m_H^2$ in Eq.~(\ref{G3H-F}) is clearly identifiable as arising
from the photon propagator and the leptonic current, and, so, it is not
necessary to treat that factor within the framework of NRQCD. We choose
not to apply the nonrelativistic expansion of NRQCD to the factor
$1/m_H^2$. That is, we apply NRQCD only to the heavy-quark factor in the
amplitude. This choice reduces the theoretical uncertainties by making
use of the fact that the quarkonium masses are known very precisely.

The order-$\alpha_s^2$ corrections to $\Gamma[H({}^3S_1)\to e^+e^-]$
(Ref.~\cite{Beneke:1997jm,Czarnecki:1997vz}) contain a strong
dependence on the NRQCD factorization scale. If one were to include
those corrections in the expression (\ref{G3H-F}) and use it to
determine $\langle\mathcal{O}_1\rangle_H$, then 
$\langle\mathcal{O}_1\rangle_H$ would 
also contain a strong dependence on the NRQCD factorization scale. If 
one were to make use of $\langle\mathcal{O}_1\rangle_H$ in calculating 
other quarkonium decay and production processes, then the 
factorization-scale dependence would cancel only if the short-distance 
coefficients for those processes were calculated through relative order 
$\alpha_s^2$. Generally, short-distance coefficients for quarkonium
processes have not been calculated beyond relative order $\alpha_s$.
For this reason, we have chosen to omit the order-$\alpha_s^2$
corrections to the leptonic width in Eq.~(\ref{G3H-F}).

\subsection{$\bm{\Gamma[H({}^1S_0)\to \gamma\gamma]}$%
\label{sec:resum-eta}}
Employing a method analogous to that which is given in
Sec.~\ref{sec:resum-psi}, one can obtain the NRQCD factorization formula
for the relativistic corrections to the two-photon decay of a
spin-singlet $S$-wave quarkonium state $H({}^1S_0)$.

The NRQCD factorization formula for the amplitude for the decay 
$H({}^1S_0)\to \gamma\gamma$ is
\begin{equation}
\mathcal{A}[H({}^1S_0)\to\gamma\gamma]
=
\sqrt{2m_H}
\sum_n d_n({}^1S_0) \langle 0|\mathcal{O}_n|H({}^1S_0)\rangle,
\label{A1H-full}%
\end{equation}
where the $d_n({}^1S_0)$ are short-distance coefficients. We
approximate this expression by keeping only those matrix elements that
connect the vacuum to the color-singlet $Q\bar{Q}$ Fock state in
the quarkonium. Then, we have 
\begin{equation}
\mathcal{A}[H({}^1S_0)\to\gamma\gamma]
=
\sqrt{2m_H}
\sum_n c_n({}^1S_0) \langle 0|
\chi^\dagger 
(-\tfrac{i}{2}\tensor{\bm{D}})^{2n} 
\psi|H({}^1S_0)\rangle.
\label{A1H}%
\end{equation}
As in the spin-triplet case, this modified NRQCD factorization formula
retains all of the relativistic corrections that are contained in a
potential-model $Q\bar Q$ wave function, up to the ultraviolet 
cutoff of the NRQCD matrix elements.

We can calculate the short-distance coefficients $c_n({}^1S_0)$ by
replacing the initial hadronic state $\sqrt{2m_H}|H({}^1S_0)\rangle$ in
Eq.~(\ref{A1H}) with a perturbative spin-singlet $S$-wave $Q\bar{Q}$
state:
\begin{eqnarray}
\mathcal{A}[Q\bar{Q}_1({}^1S_0)\to\gamma\gamma]
=
\sum_n
c_n({}^1S_0)
\langle 0|\chi^\dagger 
(-\tfrac{i}{2}\tensor{\bm{D}})^{2n}
\psi|Q\bar{Q}_1({}^1S_0)\rangle.
\label{A1Q}%
\end{eqnarray}
In the rest frame of the quarkonium, the
perturbative amplitude on the left side of Eq.~(\ref{A1Q}) at order
$\alpha_s^{0}$ is given by~\cite{Bodwin:2002hg}
\begin{equation}
\mathcal{A}[Q\bar{Q}_1({}^1S_0)\to\gamma\gamma]
=
\sqrt{2N_c}\,
 {e^2 e_Q^2}
\, 
\frac{\bm{k}_1\cdot \bm{\epsilon}_1^*\times \bm{\epsilon}_2^*
}{|\bm{k}_1|}
\,
\frac{m_Q}{|\bm{q}|}
\log\frac{E(q)+|\bm{q}|}{E(q)-|\bm{q}|},
\label{A1Q-F}%
\end{equation}
where $k_i$ and $\epsilon_i$ are the momentum and the polarization of
the $i$-th photon.
The perturbative NRQCD matrix elements on the right side of
Eq.~(\ref{A1Q}) are given by
\begin{equation}
\langle 0|\chi^\dagger 
(-\tfrac{i}{2}\tensor{\bm{D}})^{2n}
\psi|Q\bar{Q}_1({}^1S_0)\rangle
=
\sqrt{2N_c}\,2E(q)\, \bm{q}^{2n}.
\label{Qnorm1}%
\end{equation}
By comparing Eqs.~(\ref{A1Q}) and (\ref{A1Q-F}),
one can read off the short-distance coefficients
$c_n({}^1S_0)$:
\begin{equation}
c_n({}^1S_0) = 
 {e^2 e_Q^2}
\, 
\frac{\bm{k}_1\cdot \bm{\epsilon}_1^*\times \bm{\epsilon}_2^*
}{|\bm{k}_1|}
\,
\left[
\frac{1}{n!}\left(\frac{\partial }{\partial \bm{q}^2}
\right)^n
\frac{m_Q}{2E(q)|\bm{q}|}
\log\frac{E(q)+|\bm{q}|}{E(q)-|\bm{q}|}
\right]_{\bm{q}^2=0}.
\label{c1n}%
\end{equation}

Substituting the $c_n({}^1S_0)$ in Eq.~(\ref{c1n}) into Eq.~(\ref{A1H}), using
Eq.~(\ref{g-k-2}), and including the order-$\alpha_s$ correction to 
the amplitude~\cite{Barbieri:1979be,Hagiwara:1980nv,harris-brown}, we obtain
\begin{eqnarray}
\mathcal{A}[H({}^1S_0)\to\gamma\gamma]
&=& \frac{\sqrt{2m_H}}{m_Q} {e^2 e_Q^2}
\, 
\frac{\bm{k}_1\cdot \bm{\epsilon}_1^*\times \bm{\epsilon}_2^*
}{|\bm{k}_1|}
\,
\left[1-g\left(\langle \bm{q}^2\rangle_H/m_Q^2\right)
      -\frac{20-\pi^2}{8}C_F\frac{\alpha_s}{\pi}\right]
\nonumber\\
&&\times
\langle 0|\chi^\dagger\psi|H({}^1S_0)\rangle,
\label{A1H-F}%
\end{eqnarray}
where $g(x)$ is defined by
\begin{eqnarray}
g(x)&=&1-\frac{1}{2\sqrt{x(1+x)}}
\log\left[\frac{\sqrt{1+x}+\sqrt{x}}{\sqrt{1+x}-\sqrt{x}}\right]
\nonumber \\
&=& 
1-\frac{1}{2\sqrt{x(1+x)}}
\log\left[1+2\sqrt{x(1+x)}+2x\right].
\label{gx}%
\end{eqnarray}
Squaring the amplitude (\ref{A1H-F}), summing over the photon
polarizations, multiplying by the phase space and $1/2!$ for the two
identical particles in the final state, and dividing by the
normalization $2m_H$, we obtain the two-photon decay width of a
spin-singlet $S$-wave heavy quarkonium:
\begin{equation}
\Gamma[H({}^1S_0)\to\gamma\gamma]
=
\frac{2\pi \alpha^2 e_Q^4}{m_Q^2}\,
\left[1-g(\langle \bm{q}^2\rangle_H/m_Q^2)
-\frac{20-\pi^2}{8}C_F \frac{\alpha_s}{\pi}
\right]^2
\langle \mathcal{O}_1\rangle_H.
\label{G1H-F}%
\end{equation}

In the formula (\ref{G1H-F}), we have omitted the order-$\alpha_s^2$
corrections to the decay amplitude \cite{Czarnecki:2001zc}. As we
discussed in the case of the leptonic width of a spin-triplet $S$-wave
quarkonium, the order-$\alpha_s^2$ corrections contain a dependence on
the NRQCD factorization scale and can only be used consistently in
conjunction with calculations of other quarkonium processes through
order $\alpha_s^2$.

\section{Potential Model\label{sec:cornell}}
As we have explained earlier, in order to compute the higher-order
matrix elements that appear in Eq.~(\ref{me-ratios}), we need to compute
the ground-state binding energy $\epsilon_{1S}$ that appears in the
generalized Gremm-Kapustin relation (\ref{gremm-kapustin}). In this
section, we describe briefly the potential model that we use to compute
$\epsilon_{1S}$. For details of the model, we refer the reader to
Refs.~\cite{Eichten:1978tg,Bodwin:2006dn}.

The model makes use of the Cornell potential~\cite{Eichten:1978tg}, 
which parametrizes the $Q\bar Q$ potential as a linear combination of the 
Coulomb and linear potentials:
\begin{equation}
V(r)=-\frac{\kappa}{r}+\sigma r,
\label{model-V}%
\end{equation}
where $\kappa$ is a dimensionless model parameter for the Coulomb 
strength and $\sigma$ is the string tension, which is of mass dimension two.
In the original formulation of the Cornell potential model 
\cite{Eichten:1978tg}, the strength of the linear potential was given in 
terms of a parameter $a$, where
\begin{equation}
a=1/\sqrt{\sigma}.
\end{equation}
By varying the parameters in the Cornell potential, one can obtain good 
fits to lattice measurements of the $Q\bar Q$ static potential 
\cite{Bali:2000gf}. Therefore, we assume that the use of the Cornell 
parametrization of the $Q\bar Q$ potential results in errors that are 
much less than the order-$v^2$ errors (about 30\%) that are inherent in 
the leading-potential approximation to NRQCD. 

The Schr\"{o}dinger equation for the radial wave function 
$R_{n\ell}(r)$ with the radial and orbital angular-momentum
quantum numbers $n$ and $\ell$ is 
\begin{equation}
\left[
-\frac{1}{mr^2} \frac{d}{dr}\left( r^2 \frac{d}{dr}\right)
+\frac{\ell(\ell+1)}{m r^2}
+V(r)
\right]R_{n\ell}(r)=\epsilon_{n\ell} R_{n\ell}(r),
\label{radial}%
\end{equation}
where $m$ is the quark mass and $\epsilon_{n\ell}$ is the binding energy
of the $n\ell$ state. We treat $m$ as a parameter of the potential model
and note that it is, in general, different from the heavy-quark mass
$m_Q$, which appears in the short-distance coefficients of NRQCD
factorization formulas. As usual, for an $S$-wave state, the wave
function is $\psi_{nS}(r)=R_{nS}(r)/\sqrt{4\pi}$.

Introducing the scaled radius $\rho$ and scaled coupling
$\lambda$~\cite{Eichten:1978tg},
\begin{subequations}
\begin{eqnarray}
\rho&=&(\sigma m)^{1/3}\, r,
\label{r-rho}%
\\
\lambda&=&\frac{\kappa}{(\sigma/m^2)^{1/3}},
\label{kappa-lambda}%
\end{eqnarray}
\label{dimensionless}%
\end{subequations}
which are dimensionless, one can rewrite the radial equation 
(\ref{radial}) as~\cite{Eichten:1978tg}
\begin{equation}
\left[
\frac{d^2}{d\rho^2}-\frac{\ell(\ell+1)}{\rho^2} 
+\frac{\lambda}{\rho}-\rho+\zeta_{n\ell}
\right]u_{n\ell}(\rho)=0,
\label{eq:u}%
\end{equation}
where $u_{n\ell}(\rho)$ and $\zeta_{n\ell}$ are the dimensionless radial
wave function and the dimensionless energy eigenvalue of the $n\ell$
state. The relation between $R_{n\ell}(r)$ and $u_{n\ell}(\rho)$ is
\begin{equation}
\label{eq:psir}%
R_{n\ell}(r)=
\sqrt{\sigma m}
\, \frac{u_{n\ell}(\rho)}{\rho},
\end{equation}
where the wave functions are normalized according to
\begin{equation}
\int_0^\infty |u_{n\ell}(\rho)|^2 d\rho
=\int_0^\infty |R_{n\ell}(r)|^2 r^2dr=1.
\end{equation}
The binding energy is related to the dimensionless eigenvalue 
$\zeta_{n\ell}$ as
\begin{equation}
\epsilon_{n\ell}= 
[\sigma^2/m]^{1/3}\zeta_{n\ell}(\lambda).
\label{eq:ezeta}%
\end{equation}

Now let us specialize to the $S$-wave case. In order to compute
$\epsilon_{nS}$ from Eq.~(\ref{eq:ezeta}), we must fix the model
parameters $\sigma$, $m$, and $\lambda$ and solve Eq.~(\ref{eq:u}), with
$\ell=0$, for $\zeta_{nS}(\lambda)$. Our strategy is to fix $\sigma$ from
lattice measurements and to use the measured $1S$-$2S$ mass
splitting and $|\psi_{nS}(0)|^2$, as determined from the electromagnetic
decay widths, to solve for $m$ and $\lambda$.
Using Eq.~(\ref{eq:ezeta}), we can express $m$ in terms of the $1S$-$2S$ 
mass splitting:
\begin{equation}
m (\lambda)=
\sigma^2 \left[\frac{\zeta_{2S}(\lambda)-\zeta_{1S}(\lambda)}
                    {m_{2S}-m_{1S}}\right]^3,
\label{m-lam}%
\end{equation}
For $S$-wave states, the wave function at the origin
$\psi_{nS}(0)=R_{nS}(0)/\sqrt{4\pi}$ can be expressed
as~\cite{Eichten:1978tg}
\begin{equation}
|\psi_{nS}(0)|^2
=\frac{m}{4\pi}\int d^3r
   |\psi_{nS}(r)|^2
   \frac{\partial V(r)}{\partial\, r} 
=\frac{\sigma\, m(\lambda)}{4\pi}
\left[
1+\lambda F_{nS}(\lambda)
\right]
,
\label{eq:psi0rhom2}%
\end{equation}
where $F_{nS}(\lambda)$ is the expectation value of $1/\rho^2$ 
for the $nS$ state:
\begin{equation}
F_{nS}(\lambda)=
\int_0^\infty \frac{d\rho}{\rho^2} \,\left|u_{nS}(\rho)\right|^2.
\label{eq:rhom2}%
\end{equation}
The first equality in Eq.~(\ref{eq:psi0rhom2}) can be obtained by
multiplying the radial Schr\"odinger Equation (\ref{radial}) on the left
by $R_{nS}^*(r)$ and integrating by parts.

For purposes of computation of the NRQCD matrix elements, it is
convenient to express those matrix elements in terms of the
potential-model parameters. From Eqs.~(\ref{eq:ezeta}) and 
(\ref{eq:psi0rhom2})
and the generalized Gremm-Kapustin relation
(\ref{gremm-kapustin}), we find that
\begin{subequations}
\label{me-params}%
\begin{eqnarray}
\langle \mathcal{O}_1\rangle_{H}
&=&
2N_c
|\psi(0)|^2
=
\frac{\sigma N_c\, m(\lambda)}{2\pi}
\left[
1+\lambda F_{1S}(\lambda)
\right],
\label{O-params}%
\\
\langle \bm{q}^2\rangle_{H}
&=& m(\lambda)\epsilon_{1S}(\lambda)
=[\sigma m(\lambda)]^{2/3}\zeta_{1S}(\lambda),
\label{psq-params}%
\end{eqnarray}
\end{subequations}
where $m(\lambda)$ is given in Eq.~(\ref{m-lam}) and $F_{1S}(\lambda)$ is 
given in Eq.~(\ref{eq:rhom2}).

\section{Computation of the NRQCD matrix elements\label{sec:det}}
In this section, we determine the numerical values of the NRQCD matrix
elements for the $J/\psi$ and the $\eta_c$. In this and
subsequent discussions, we drop the superscript $\textrm{VS}$ on
$\langle\mathcal{O}_1\rangle_H^\textrm{VS}$ because other sources of
uncertainty, which we will describe, are much larger than the error in
the vacuum-saturation approximation.

\subsection{Method of computation}

Were it not for the relativistic corrections in the decay widths in
Eqs.~(\ref{G3H-F}) and (\ref{G1H-F}), we could simply solve those
equations for $\langle \mathcal{O}_1\rangle_H$. Then we could use
the value for $\langle \mathcal{O}_1\rangle_H$ that we would obtain to
solve Eq.~(\ref{O-params}) for $\lambda$ and use that value of
$\lambda$ to solve Eq.~(\ref{psq-params}) for $\langle
\bm{q}^2\rangle_H$ . Because the relativistic corrections in
Eqs.~(\ref{G3H-F}) and (\ref{G1H-F}) couple those equations weakly to
Eq.~(\ref{psq-params}), we must carry out the more difficult task of
solving Eq.~(\ref{G3H-F}) or Eq.~(\ref{G1H-F}) simultaneously with
Eqs.~(\ref{O-params}) and (\ref{psq-params}).

First, we express $\Gamma[J/\psi\to e^+e^-]$ and
$\Gamma[\eta_c\to\gamma\gamma]$ in terms of the potential-model
parameters by substituting Eq.~(\ref{me-params}) into Eq.~(\ref{G3H-F}) and
Eq.~(\ref{G1H-F}), respectively. We equate those expressions to the 
experimental values of the electromagnetic widths \cite{Yao:2006px}:
\begin{subequations}
\label{exp-widths}%
\begin{eqnarray}
\Gamma[J/\psi\to e^+e^-]&=&5.55\pm 0.14\pm 0.02~\hbox{keV},
\label{psi-exp}%
\\
\Gamma[\eta_c\to\gamma\gamma]&=&7.2\pm 0.7\pm 2.0~\hbox{keV}.
\label{etac-exp}%
\end{eqnarray}
\end{subequations}
Then, we solve the resulting equations numerically for the model 
parameter $\lambda$. In computing the solution, we express the 
eigenvalues $\zeta_{1S}(\lambda)$ and $\zeta_{2S}(\lambda)$ and the 
expectation value $F_{1S}(\lambda)$ [Eq.~(\ref{eq:rhom2})] as functions 
of $\lambda$ by fitting interpolating polynomials to computations of 
the eigenvalues and expectation value at fixed values of $\lambda$. Once 
we have obtained a value for $\lambda$, we substitute it into 
Eq.~(\ref{me-params}) to obtain values for the NRQCD matrix elements.

In carrying out the numerical computation, we need values for 
the charm-quark mass $m_c$, the string tension $\sigma$, and the $1S$-$2S$ mass 
splitting. In order to maintain consistency with the calculations of
the electromagnetic decay widths of the $J/\psi$ and the $\eta_c$ at NLO
in $\alpha_s$, we take $m_c$ to be the pole mass. The specific
numerical value that we use is\footnote{The most recent compilation of
the Particle Data Group \cite{Yao:2006px} suggests that the actual      
uncertainty in $m_c$ may be a factor of two smaller than the uncertainty
that we use here. However, since it is not clear that the
systematic errors are well understood in the various determinations that
enter into that compilation, we make a conservative choice of error bars.}
\begin{equation}
m_c=1.4\pm 0.2~\hbox{GeV}.
\label{quark-mass}%
\end{equation}

We fix the string tension $\sigma$ by making use of lattice
measurements. From Ref.~\cite{Booth:1992bm}, we find that $\sigma
a_{\rm L}^2=0.0114(2)$ at a lattice coupling $\beta=6.5$, where $a_{\rm
L}$ is the lattice spacing. Lattice calculations of the hadron spectrum
at $\beta=6.5$ yield values for $1/a_{\rm L}$ of $3.962(127)$~GeV  
(Refs.~\cite{Gupta:1996sa,Kim:1993gc}) and $3.811(59)$~GeV
(Refs.~\cite{Gupta:1996sa,Kim:1996cz}). These result in values for the
string tension of $\sigma=0.1790\pm 0.0119$~GeV$^2$ and
$\sigma=0.1656\pm 0.0059$~GeV$^2$, respectively.
Combining these two values, we obtain
\begin{equation}
\sigma=0.1682\pm 0.0053~\textrm{GeV}^2.
\label{eq:sigma}%
\end{equation}
For the $1S$-$2S$ mass splitting, we take the mass difference between
the $J/\psi$ and $\psi(2S)$ \cite{Yao:2006px}:
\begin{equation}
m_{2S}-m_{1S}=589.177\pm 0.036~\hbox{MeV}.
\end{equation}
We use $m_{J/\psi}=3.096916$~GeV and $m_{\eta_c}=2.9798$~GeV           
\cite{Yao:2006px}.
We also need values for $\alpha_s$. In the case of $J/\psi \to e^+
e^-$, we choose the scale of $\alpha_s$ to be that of the momentum
transfer at the virtual-photon-charm-quark vertex, namely, $m_{J/\psi}$.
In the case of $\eta_c\to \gamma\gamma$, we choose the scale of
$\alpha_s$ to be that of the momentum transfer at either of the
photon-charm-quark vertices, namely, $m_{\eta_c}/2$.\footnote{We compute
$\alpha_s$ and $\alpha$ at each scale by making use of the code GLOBAL
ANALYSIS OF PARTICLE PROPERTIES (GAPP)~\cite{Erler:1998sy}.}  In order
to take into account uncertainties in the scale and omitted corrections
to the decay rates of next-to-next-to-leading order (NNLO) in
$\alpha_s$, we attach an uncertainty to $\alpha_s$ whose relative size
is $\alpha_s$. Then, we have
\begin{subequations}
\label{alphas}%
\begin{eqnarray}
\label{alphasJpsi}%
\alpha_s(m_{J/\psi})&=&0.25\pm 0.06,
\\
\label{alphasetac}%
\alpha_s(m_{\eta_c}/2)&=&0.35\pm 0.12.
\end{eqnarray}
\end{subequations}
We choose the scales for the running QED coupling 
$\alpha$ to be the same as those for $\alpha_s$:
\begin{subequations}
\label{alpha}%
\begin{eqnarray}
\label{alphaJpsi}%
\alpha(m_{J/\psi})&=&\frac{1}{132.6},
\\
\label{alphaetac}%
\alpha(m_{\eta_c}/2)&=&\frac{1}{133.6},
\end{eqnarray}
\end{subequations}
where we ignore the uncertainties in $\alpha$.

In the case of the $\eta_c$ matrix elements, we actually make use of two
methods of computation. One method is to compute the $\eta_c$ matrix
elements from $\Gamma[\eta_c\to\gamma\gamma]$, as we have outlined
above. A second method is to equate the $\eta_c$ matrix elements to the
$J/\psi$ matrix elements that we determine from $\Gamma[J/\psi\to
e^+e^-]$. Owing to the approximate heavy-quark spin symmetry 
of NRQCD \cite{Bodwin:1994jh}, this
equality is valid up to corrections of relative order $v^2$. By
combining these two methods of determining the $\eta_c$ matrix elements,
we can reduce the uncertainties. This approach is useful because the
experimental result for $\Gamma[\eta_c\to\gamma\gamma]$
[Eq.~(\ref{etac-exp})] has a relative uncertainty that is comparable to
the corrections to the spin-symmetry relation, which are of order
$v^2\approx 30\%$. In principle, we could apply a similar approach to
the $J/\psi$ matrix elements, but we would not gain a significant
reduction in the uncertainties because the relative uncertainty in the
experimental result for $\Gamma[J/\psi \to e^+e^-]$
[Eq.~(\ref{psi-exp})] is small compared to the corrections to the
spin-symmetry relation. In averaging the two sets of $\eta_c$ matrix
elements, we must take into account the fact that many of the
uncertainties are correlated between the two sets. We describe the
procedure that we use for doing this in detail in the next section.

\subsection{Sources of uncertainties\label{sec:num}} 
Let us now list the various uncertainties that enter into the
calculations of the matrix elements. There is a theoretical uncertainty
in the value of $\langle \bm{q}^2\rangle_H$ that arises from the fact
that the leading-potential approximation is accurate only up to
corrections of relative order $v^2$.
 For the computation that is based
on $\Gamma[J/\psi\to e^+e^-]$, we denote this uncertainty by
$\Delta\langle \bm{q}^2\rangle_{J/\psi}$, and for the computation that
is based on $\Gamma[\eta_c\to \gamma\gamma]$, we denote this uncertainty
by $\Delta\langle \bm{q}^2\rangle_{\eta_c}$. We take these
uncertainties to be $v^2\approx 30\%$ times the central values. The
uncertainties that arise from the scale uncertainties in $\alpha_s$ and
from neglecting NNLO corrections to the $J/\psi$ and $\eta_c$
electromagnetic widths are denoted by $\Delta{\rm NNLO}_{J/\psi}$,
$\Delta{\rm NNLO}_{\eta_c}$, respectively. As we have explained
above, we parametrize these uncertainties as uncertainties in $\alpha_s$
[Eq.~(\ref{alphas})]. However, we take $\Delta{\rm NNLO}_{J/\psi}$ and
$\Delta{\rm NNLO}_{\eta_c}$, to be uncorrelated. There are also
uncertainties that are associated with the charm-quark mass $m_c$
[Eq.~(\ref{quark-mass})], the string tension $\sigma$
[Eq.~(\ref{eq:sigma})], and the uncertainties in the experimental
measurements of $\Gamma[J/\psi\to e^+e^-]$ and $\Gamma[\eta_c\to
\gamma\gamma]$ [Eqs.~(\ref{psi-exp}) and (\ref{etac-exp})]. We denote
these uncertainties by $\Delta m_c$, $\Delta\sigma$,
$\Delta\Gamma_{J/\psi}$, and $\Delta\Gamma_{\eta_c}$, respectively. When
we combine the values of the $\eta_c$ matrix elements that we obtain from
$\Gamma[\eta_c\to \gamma\gamma]$ with those that we obtain from
$\Gamma[J/\psi\to e^+e^-]$ by invoking the heavy-quark spin symmetry,
there is an uncertainty from corrections to the spin symmetry, which
applies to the latter set of matrix elements. We take it to be
$v^2\approx 30\%$ times the values of that set of matrix elements. Since
$\langle\bm{q}^2\rangle_{J/\psi}$ already has an uncertainty
$\Delta\langle\bm{q}^2\rangle_{J/\psi}$ of order $v^2$, we apply this
additional order-$v^2$ uncertainty only to
$\langle\mathcal{O}_1\rangle_{J/\psi}$. We denote it by $\Delta v^2$.

In making these uncertainty estimates, we have assumed that the
standard NRQCD power-counting (velocity-scaling) rules
\cite{Bodwin:1994jh} hold. Various alternatives to the NRQCD
power-counting rules have been suggested
\cite{Brambilla:1999xf,Pineda:2000sz,Fleming:2000ib}. Application of
these alternative rules would affect our estimate of the correction to
the heavy-quark spin symmetry, $\Delta v^2$, and our estimates of the
corrections to the static potential, $\Delta\langle
\bm{q}^2\rangle_{J/\psi}$ and $\Delta\langle \bm{q}^2\rangle_{\eta_c}$.
In the standard NRQCD power-counting rules, $\Delta v^2$ is of relative order
$v^2$.  In the strong-coupling regime of 
Refs.~\cite{Brambilla:1999xf,Pineda:2000sz}
and in the power-counting rules of Ref.~\cite{Fleming:2000ib}, $\Delta
v^2$ is of relative order $\Lambda_{\textrm{QCD}}/m_c$, which is actually smaller
numerically than $v^2$. The leading correction to the static potential
is denoted by $V^{(1)}/m$ (Ref.~\cite{Brambilla:1999xf}). In the
standard NRQCD power-counting rules, $V^{(1)}/m$ is suppressed as
$v^2$ relative to the static potential. In the strong-coupling regime of
Refs.~\cite{Brambilla:1999xf,Pineda:2000sz} and in the power-counting rules of
Ref.~\cite{Fleming:2000ib}, $V^{(1)}/m$ is of the same order as the
static potential \cite{Pineda:2000sz,Brambilla:2004jw}. In the lattice calculation of
Ref.~\cite{Koma:2007jq}, $V^{(1)}/m$ corrects the string tension by about
17\%, which is numerically smaller than $v^2$. Other lattice
calculations \cite{bks,Bodwin:2005gg,Bodwin:2004up} also suggest that
terms of higher order in the standard NRQCD power counting are
suppressed at least as much as would be expected from the standard power
counting. Therefore, we believe that the standard NRQCD power-counting
rules give an upper bound on the uncertainties, and we use them for our
uncertainty estimates. One could implement the alternative power-counting
rules by equating $\Delta v^2$ to $\Lambda_\textrm{QCD}/m_c$ times
the central value and by
by equating $\Delta \langle \bm{q}^2 \rangle_{J/\psi}$ and 
$\Delta \langle \bm{q}^2 \rangle_{\eta_c}$ to $100\%$ of the central value.

\subsection{Numerical results}
\subsubsection{Computations using $\Gamma[J/\psi\to e^+e^-]$ and 
$\Gamma[\eta_c\to \gamma\gamma]$\label{basic-comps}}

The results of our computations of matrix elements from
$\Gamma[J/\psi\to e^+e^-]$ and  $\Gamma[\eta_c\to \gamma\gamma]$ are
shown in Tables~\ref{table1} and \ref{table2}, respectively. In each
table, in the first row below the headings, we give the central values
for the potential-model parameter $\lambda$, the matrix element
$\langle\mathcal{O}_1\rangle_{H}$, and the ratio $\langle\bm{q}^2
\rangle_{H}$. Subsequent rows contain the
values for $\lambda$, the matrix element, and the ratio that result from
shifting each uncertain quantity in the calculation by plus or minus its
uncertainty. We put a superscript $\gamma\gamma$ on the matrix element
and the ratio for the $\eta_c$ that are shown in Table~\ref{table2},
in order to specify that these numbers are the result of a fit to
$\Gamma[\eta_c\to\gamma\gamma]$.

The matrix elements and ratios of matrix elements, along with their
uncertainties, are as follows:
\begin{subequations}
\label{psi-me-nums}%
\begin{eqnarray}
\langle \mathcal{O}_1 \rangle_{J/\psi}&=&
0.440^{+0.009+0.011+0.003+0.064+0.011}
     _{-0.010-0.008-0.003-0.053-0.011}~\textrm{GeV}^3
=0.440^{+0.067}_{-0.055}~\textrm{GeV}^3,
\\
\langle \bm{q}^2 \rangle_{J/\psi}&=&
0.441^{+0.132+0.003+0.041+0.018+0.004}
     _{-0.132-0.004-0.040-0.022-0.004}~\textrm{GeV}^2
=0.441^{+0.140}_{-0.140}~\textrm{GeV}^2,
\end{eqnarray}
\end{subequations}
\begin{subequations}
\begin{eqnarray}
\langle \mathcal{O}_1 \rangle_{\eta_c}^{\gamma\gamma}&=&
0.434^{+0.042+0.083+0.012+0.083+0.112}
     _{-0.040-0.069-0.012-0.066-0.118}~\textrm{GeV}^3
=0.434^{+0.169}_{-0.158}~\textrm{GeV}^3,
\\
\langle \bm{q}^2 \rangle_{\eta_c}^{\gamma\gamma}&=&
0.443^{+0.133+0.024+0.038+0.023+0.041}
     _{-0.133-0.028-0.037-0.028-0.038}~\textrm{GeV}^2
=0.443^{+0.148}_{-0.149}~\textrm{GeV}^2.
\end{eqnarray}
\label{etac-gam-gam-me-nums}%
\end{subequations}
In the first equalities in Eqs.~(\ref{psi-me-nums}) and
(\ref{etac-gam-gam-me-nums}), the uncertainties are presented in the
same order as in Tables~\ref{table1} and \ref{table2}. In the last
equalities in each of these equations, we have added the uncertainties in
quadrature. However it must be kept in mind for many applications that
the individual uncertainties are correlated between the matrix elements.
The correlations can be determined from the tabulations in
Tables~\ref{table1}  and \ref{table2}.

From Eqs.~(\ref{v-sq}), (\ref{psi-me-nums}), and
(\ref{etac-gam-gam-me-nums}) and the uncertainty in $m_c$ in
Eq.~(\ref{quark-mass}), it can be deduced that
\begin{subequations}
\label{vsq}
\begin{eqnarray}
\langle v^2 \rangle_{J/\psi}&=&
0.225^{+0.106}_{-0.088},\\
\langle v^2 \rangle_{\eta_c}^{\gamma\gamma}&=&
0.226^{+0.123}_{-0.098}.
\end{eqnarray}
\end{subequations}
The central values of these results are somewhat smaller than an
estimate, based on the NRQCD velocity-scaling rules
\cite{Bodwin:1994jh}, that $\langle v^2 \rangle$ should be equal
approximately to $v^2\approx 0.3$. However, they are consistent with
being of order $v^2$.

We can see the effect of resummation by repeating our analysis, but
keeping only the order-$v^2$ corrections in the formulas for the decay
rates in Eqs.~(\ref{G3H-F}) and (\ref{G1H-F}). The results are that the
central values are shifted to $\langle
\mathcal{O}_1\rangle_{J/\psi}=0.446269$~GeV$^3$, $\langle\bm{q}^2
\rangle_{J/\psi}=0.438520$~GeV$^2$, $\langle
\mathcal{O}_1\rangle_{\eta_c}^{\gamma\gamma}=0.459867$~GeV$^3$, and
$\langle\bm{q}^2 \rangle_{\eta_c}^{\gamma\gamma}=0.433879$~GeV$^2$.
Hence, the effects of the resummation on these quantities are $-1.4\%$,
$+0.5\%$, $-5.7\%$, and $+2.1\%$, respectively. The small effects from
resummation suggest that the $v$ expansion of NRQCD converges well for
the widths $\Gamma[J/\psi\to e^+e^-]$ and $\Gamma[\eta_c\to\gamma\gamma]$.

\subsubsection{Average values of $\eta_c$ matrix elements}
 
Because some of the uncertainties in Tables~\ref{table1} and
\ref{table2} are correlated, we must take care in combining the results
in these tables to obtain average values for the $\eta_c$ matrix element
and the $\eta_c$ ratio of matrix elements. First, we construct a
two-by-two covariance matrix for the quantities 
$\langle \mathcal{O}_1\rangle_{J/\psi}$ and
$\langle \mathcal{O}_1\rangle_{\eta_c}^{\gamma\gamma}$ from the
deviations from the central values that correspond to the uncertainties
listed in Tables~\ref{table1} and \ref{table2}. Then, we use the inverse
of the covariance matrix to construct $\chi^2$ for the deviation of
the average value of $\langle \mathcal{O}_1\rangle_{\eta_c}$ from the
two input values. We fix the average value of
$\langle \mathcal{O}_1\rangle_{\eta_c}$  by minimizing this $\chi^2$
with respect to it. The minimum value of $\chi^2$ is $8.9\times
10^{-4}$. This small value of $\chi^2$ reflects the fact that $\langle
\mathcal{O}_1\rangle_{J/\psi}$ and $\langle
\mathcal{O}_1\rangle_{\eta_c}^{\gamma\gamma}$ are much closer in value
than one would expect from the velocity scaling rules of NRQCD. Once
we have obtained the average value of $\langle
\mathcal{O}_1\rangle_{\eta_c}$, we use it as an input to the potential
model to compute the average value of $\langle\bm{q}^2
\rangle_{\eta_c}$. We carry out this computation of the average values
of $\langle \mathcal{O}_1\rangle_{\eta_c}$ and $\langle\bm{q}^2
\rangle_{\eta_c}$ for values of the input parameters that correspond to
each of the uncertainties that we have described. (The effect of the
uncertainty $\Delta \langle\bm{q}^2\rangle_{\eta_c}$ on the average
value of $\langle \mathcal{O}_1\rangle_{\eta_c}$ has already been taken
into account through the inputs to that average. We obtain the effect of
$\Delta \langle\bm{q}^2\rangle_{\eta_c}$ on the average value of
$\langle\bm{q}^2\rangle_{\eta_c}$ by varying the central value of the
average value of $\langle\bm{q}^2\rangle_{\eta_c}$ by $v^2\approx
30\%$.) The average values of $\langle
\mathcal{O}_1\rangle_{\eta_c}$ and $\langle\bm{q}^2 \rangle_{\eta_c}$
that result from these computations are shown in Table~\ref{table3}.
The first row after the headings in Table~\ref{table3} gives the central
values of the averages of the $\eta_c$ matrix element
$\langle\mathcal{O}_1\rangle_{\eta_c}$ and ratio of $\eta_c$ matrix
elements $\langle\bm{q}^2 \rangle_{\eta_c}$. Subsequent rows show the
effects of the various uncertainties on the average values. The central
values and uncertainties in Table~\ref{table3} can be summarized as
follows:
\begin{subequations}
\label{etac-me-with-mc}%
\begin{eqnarray}
\langle \mathcal{O}_1 \rangle_{\eta_c}&=&
0.437^{+0.024 +0.033 +0.007 +0.036 +0.006 +0.073 +0.037 +0.050}
     _{-0.023 -0.025 -0.007 -0.029 -0.006 -0.073 -0.029 -0.053}
~\textrm{GeV}^3
   =0.437^{+0.111}_{-0.105}~\textrm{GeV}^3,\quad
\\
\langle \bm{q}^2 \rangle_{\eta_c}&=&
0.442^{+0.132 +0.009 +0.040 +0.010 +0.002 +0.026 +0.010 +0.018}
     _{-0.132 -0.011 -0.039 -0.012 -0.002 -0.025 -0.013 -0.017}
~\textrm{GeV}^2
=0.442^{+0.143}_{-0.143}~\textrm{GeV}^2.\quad
\end{eqnarray}
\end{subequations}
In the first equalities in Eq.~(\ref{etac-me-with-mc}), the
uncertainties are presented in the same order as in Table~\ref{table3}.
In the last equalities, we have added the uncertainties in quadrature.
As we have mentioned, it must be kept in mind for many applications that
the individual uncertainties are correlated between the matrix elements.
The correlations can be determined from the tabulations in
Table~\ref{table3}.

The correlated errors can also be expressed conveniently in terms of a 
correlation matrix. We construct a (symmetric) correlation matrix 
whose rows and 
columns correspond to $\langle \mathcal{O}_1\rangle_{J/\psi}$,
$\langle\bm{q}^2 \rangle_{J/\psi}$, 
$\langle \mathcal{O}_1\rangle_{\eta_c}$,
and $\langle\bm{q}^2 \rangle_{\eta_c}$, respectively, taking the 
deviations from the central values from Table~\ref{table1} and 
Table~\ref{table3}. The result is
\begin{equation}
C_1 = \left(
\begin{array}{cccc}
  \phantom{-}3.71\times10^{-3}& \phantom{-}1.64\times10^{-4} 
   & \phantom{-}1.94\times10^{-3} & \phantom{-}8.18\times10^{-4}\\
  \phantom{-}1.64\times10^{-4}& \phantom{-}1.96\times10^{-2} 
   &\phantom{-}2.84\times10^{-3} & \phantom{-}1.93\times10^{-2}\\
  \phantom{-}1.94\times10^{-3}&\phantom{-}2.84\times10^{-3} 
   & \phantom{-}1.16\times10^{-2} &-3.71\times10^{-4}\\
  \phantom{-}8.18\times10^{-4}& \phantom{-}1.93\times10^{-2} 
   &-3.71\times10^{-4} & \phantom{-}2.04\times10^{-2}
\end{array}
\right),
\label{c1}%
\end{equation}
where the quantity in $i$-th row and $j$-th column is expressed 
in units of $\textrm{GeV}^{n_i+n_j}$, with $n_1=n_3=3$ and $n_2=n_4=2$.
In charmonium decay and production processes, 
the NRQCD short-distance coefficients typically depend on $m_c$. Hence, 
there may be correlations between the matrix elements and short-distance 
coefficients with respect to the uncertainty in $m_c$. Therefore, we also 
give the correlation matrix for $\langle \mathcal{O}_1\rangle_{J/\psi}$,
$\langle\bm{q}^2 \rangle_{J/\psi}$, 
$\langle \mathcal{O}_1\rangle_{\eta_c}$,
and $\langle\bm{q}^2 \rangle_{\eta_c}$, respectively, in which we omit the 
uncertainties that arise from $m_c$:
\begin{equation}
C_2 = \left(
\begin{array}{cccc}
  \phantom{-}3.63\times10^{-3}& \phantom{-}1.93\times10^{-4} 
   & \phantom{-}2.21\times10^{-3} & \phantom{-}7.27\times10^{-4}\\
  \phantom{-}1.93\times10^{-4}& \phantom{-}1.96\times10^{-2} 
   &\phantom{-}2.75\times10^{-3} & \phantom{-}1.94\times10^{-2}\\
  \phantom{-}2.21\times10^{-3}&\phantom{-}2.75\times10^{-3} 
   & \phantom{-}1.08\times10^{-2} &-8.86\times10^{-5}\\
  \phantom{-}7.27\times10^{-4}& \phantom{-}1.94\times10^{-2} 
   &-8.86\times10^{-5} & \phantom{-}2.03\times10^{-2}
\end{array}
\right),
\end{equation}
where the dimensions of the elements of $C_2$ are the same as those of the 
corresponding elements of $C_1$ in Eq.~(\ref{c1}).
We note that both
correlation matrices $C_1$ and $C_2$ contain large off-diagonal elements
that correspond to a correlation between the uncertainty in
$\langle\bm{q}^2 \rangle_{J/\psi}$ and the uncertainty in
$\langle\bm{q}^2 \rangle_{\eta_c}$. Most of this correlation arises from
the uncertainty in the string tension $\sigma$.

\begin{table}[ht]
\caption{
\label{table1}%
The potential-model parameter $\lambda$, the NRQCD matrix element $\langle
\mathcal{O}_1\rangle_{J/\psi}$, and the ratio $\langle\bm{q}^2
\rangle_{J/\psi}$, as obtained from $\Gamma[J/\psi\to e^+e^-]$. The first
row below the headings contains the central values for $\lambda$, the
matrix element, and the ratio. Subsequent rows contain the maximum and 
minimum values for these quantities that are obtained by varying them 
with respect to each uncertainty.}
\begin{ruledtabular}
\begin{tabular}{lccc}
Case&
$\lambda$&
$\langle \mathcal{O}_1\rangle_{J/\psi}\,$(GeV$^3$) &
$\langle\bm{q}^2 \rangle_{J/\psi}\,$(GeV$^2$) 
\\
\hline
central&
1.243  &  0.440 &  0.441 \\ $+\Delta \langle \bm{q}^2\rangle_{J/\psi}$&
1.256  &  0.450 &  0.573 \\ $-\Delta \langle \bm{q}^2\rangle_{J/\psi}$&
1.230  &  0.430 &  0.308 \\ $+\Delta m_c$&
1.233  &  0.433 &  0.443 \\ $-\Delta m_c$&
1.258  &  0.451 &  0.437 \\ $+\Delta \sigma$&
1.191  &  0.443 &  0.482 \\ $-\Delta \sigma$&
1.297  &  0.437 &  0.400 \\ $+\Delta \,\textrm{NNLO}_{J/\psi}$&
1.325  &  0.504 &  0.419 \\ $-\Delta \,\textrm{NNLO}_{J/\psi}$&
1.166  &  0.387 &  0.459 \\ $+\Delta \Gamma_{J/\psi}$&
1.258  &  0.451 &  0.437 \\ $-\Delta \Gamma_{J/\psi}$&
1.228  &  0.429 &  0.444 
\end{tabular}
\end{ruledtabular}
\end{table}
\begin{table}[ht]
\caption{
\label{table2}%
The potential-model parameter $\lambda$, the NRQCD matrix element
$\langle \mathcal{O}_1\rangle_{\eta_c}$, and the ratio $\langle\bm{q}^2
\rangle_{\eta_c}$, as obtained from $\Gamma[\eta_c\to
\gamma\gamma]$. The first row below the headings contains the central
values for $\lambda$, the matrix element, and the ratio. Subsequent rows
contain the maximum and minimum values for these quantities that are
obtained by varying them with respect to each uncertainty.}
\begin{ruledtabular}
\begin{tabular}{lccc}
Case&
$\lambda$&
$\langle \mathcal{O}_1\rangle_{\eta_c}^{\gamma\gamma}\,$(GeV$^3$) &
$\langle\bm{q}^2 \rangle_{\eta_c}^{\gamma\gamma}\,$(GeV$^2$) 
\\
\hline
central&
1.234 & 0.434 & 0.443 \\ $+\Delta \langle \bm{q}^2\rangle_{\eta_c}$&
1.291 & 0.476 & 0.576 \\ $-\Delta \langle \bm{q}^2\rangle_{\eta_c}$&
1.175 & 0.393 & 0.310 \\ $+\Delta m_c$&
1.340 & 0.517 & 0.415 \\ $-\Delta m_c$&
1.129 & 0.364 & 0.467 \\ $+\Delta \sigma$&
1.195 & 0.446 & 0.481 \\ $-\Delta \sigma$&
1.276 & 0.422 & 0.406 \\ $+\Delta \,\textrm{NNLO}_{\eta_c}$&
1.340 & 0.517 & 0.415 \\ $-\Delta \,\textrm{NNLO}_{\eta_c}$&
1.134 & 0.368 & 0.466 \\ $+\Delta \Gamma_{\eta_c}$&
1.374 & 0.546 & 0.405 \\ $-\Delta \Gamma_{\eta_c}$&
1.041 & 0.315 & 0.484 
\end{tabular}
\end{ruledtabular}
\end{table}
\begin{table}[ht]
\caption{Average values of the NRQCD matrix element $\langle
\mathcal{O}_1\rangle_{\eta_c}$ and the ratio
$\langle\bm{q}^2\rangle_{\eta_c}$. The method of averaging is described
in the text. The first row below the headings contains the central
values for the matrix element and the ratio. Subsequent rows contain the
maximum and minimum values for these quantities that are obtained by
varying them with respect to each uncertainty.
\label{table3}%
}
\begin{ruledtabular}
\begin{tabular}{lcc}
Case &
$\langle \mathcal{O}_1 \rangle_{\eta_c}\,$(GeV$^3$) &
$\langle \bm{q}^2 \rangle_{\eta_c}\,$(GeV$^2$)
\\
\hline
central&
0.437 & 0.442 \\ $ +\Delta \langle \bm{q}^2\rangle_{\eta_c}$&
0.461 & 0.574 \\ $ -\Delta \langle \bm{q}^2\rangle_{\eta_c}$&
0.414 & 0.309 \\ $ +\Delta m_c$&
0.470 & 0.430 \\ $ -\Delta m_c$&
0.413 & 0.450 \\ $ +\Delta \sigma$&
0.444 & 0.482 \\ $ -\Delta \sigma$&
0.431 & 0.403 \\ $ +\Delta \,\textrm{NNLO}_{J/\psi}$&
0.473 & 0.429 \\ $ -\Delta \,\textrm{NNLO}_{J/\psi}$&
0.408 & 0.452 \\ $ +\Delta \Gamma_{J/\psi}$&
0.443 & 0.440 \\ $ -\Delta \Gamma_{J/\psi}$&
0.431 & 0.444 \\ $ +\Delta v^2$&
0.511 & 0.417 \\ $ -\Delta v^2$&
0.364 & 0.467 \\ $ +\Delta \,\textrm{NNLO}_{\eta_c}$&
0.474 & 0.429 \\ $ -\Delta \,\textrm{NNLO}_{\eta_c}$&
0.408 & 0.452 \\ $ +\Delta \Gamma_{\eta_c}$&
0.487 & 0.425 \\ $ -\Delta \Gamma_{\eta_c}$&                 
0.385 & 0.460
\end{tabular} 
\end{ruledtabular}
\end{table}

\section{Comparisons with previous calculations\label{sec:comparisons}}

Our results for the matrix elements can be compared with those in
Ref.~\cite{Braaten:2002fi}. In that paper, the values $\langle
\mathcal{O}_1\rangle_{J/\psi}^{\textrm{BL}} =0.335\pm
0.024\hbox{~GeV}^3$ and $\langle
\mathcal{O}_1\rangle_{\eta_c}^{\textrm{BL}} =0.297\pm
0.032\hbox{~GeV}^3$ are given. In the case of $\langle
\mathcal{O}_1\rangle_{J/\psi}$, our result is $31\%$ larger than that in
Ref.~\cite{Braaten:2002fi}. Approximately $6\%$ of that change is the
result of the change in the experimental value of $\Gamma[J/\psi\to
e^+e^-]$ from $5.26\pm 0.37\hbox{~keV}$~\cite{Hagiwara:2002fs} to
$5.55\pm 0.14\pm 0.02\hbox{~keV}$~\cite{Yao:2006px}. An implicit
relativistic correction of about $22\%$ arises from the use of
$m_{J/\psi}$ in Eq.~(\ref{G3H-F}), rather than $2m_c$.  The use of
$\alpha(m_{J/\psi})=1/132.6$ [Eq.~(\ref{alphaJpsi})], rather than
$\alpha=1/137$, decreases $\langle \mathcal{O}_1 \rangle_{J/\psi}$ by
approximately $6\%$. The remaining change of about $9\%$ is the result
of including the explicit relativistic corrections in Eq.~(\ref{G3H-F}).
In the case of $\langle
\mathcal{O}_1\rangle_{\eta_c}^{\gamma\gamma}$, our result in 
Eq.~(\ref{etac-gam-gam-me-nums}) is $46\%$
larger than the value $\langle
\mathcal{O}_1\rangle_{\eta_c}^{\textrm{BL}} =0.297\pm
0.032\hbox{~GeV}^3$ that is given in Ref.~\cite{Braaten:2002fi}. In
this case, there is a decrease in the value of $\langle \mathcal{O}_1
\rangle_{\eta_c}^{\gamma\gamma}$ of $4\%$, owing to the change
in the experimental value of $\Gamma[\eta_c\to\gamma \gamma]$ from
$7.5\pm 0.8\hbox{~keV}$~\cite{Hagiwara:2002fs} to $7.2\pm 0.7\pm
2.0\hbox{~keV}$~\cite{Yao:2006px}.  The use of
$\alpha(m_{\eta_c}/2)=1/133.6$ [Eq.~(\ref{alphaetac})], rather than
$\alpha=1/137$, decreases 
$\langle \mathcal{O}_1 \rangle_{\eta_c}^{\gamma\gamma}$
by approximately $5\%$. The use of $\alpha_s(m_{\eta_c}/2)=0.35$ 
[Eq.~(\ref{alphasetac})], rather than
$\alpha_s(m_{J/\psi})=0.25$ [Eq.~(\ref{alphasJpsi})], which is used 
for the process $\eta_c \to \gamma\gamma$ in Ref.~\cite{Braaten:2002fi}, 
enhances the matrix element by approximately $14\%$.
The remaining change of about $41\%$ arises from the relativistic
corrections in Eq.~(\ref{G1H-F}).

In Ref.~\cite{Braaten:2002fi}, the values
$\langle\bm{q}^2\rangle_{J/\psi}=0.43$~GeV$^2$ and
$\langle\bm{q}^2\rangle_{\eta_c}^{\gamma\gamma}=0.25$~GeV$^2$ were
obtained by making use of the Gremm-Kapustin \cite{Gremm:1997dq}
relation for the physical quarkonium mass and $m_c$. While these
results are not far from those in Eqs.~(\ref{psi-me-nums}) and
(\ref{etac-gam-gam-me-nums}), the uncertainties given in
Ref.~\cite{Braaten:2002fi} are on the order $100\%$, owing to the
uncertainty in $m_c$. In our calculation, we have been able to reduce
the uncertainties significantly by making use of the Gremm-Kapustin
relation (\ref{gremm-kapustin}) for the binding energy in the potential
model. This leads to much smaller uncertainties than the use of the
Gremm-Kapustin relation for the physical quarkonium mass and $m_c$
because we compute the binding energy directly in the potential model,
instead of expressing it as a difference between $m_H$ and $2m_c$.

In Ref.~\cite{Bodwin:2006dn}, the result
$\langle\bm{q}^2\rangle_{J/\psi}=0.50\pm 0.09\pm 0.15$~GeV$^2$ was
obtained from a potential-model calculation, which also made use of the
Cornell potential. That result agrees, within errors, with the result in
Eq.~(\ref{psi-me-nums}). In Ref.~\cite{Bodwin:2006dn}, the value of the
matrix element $\langle \mathcal{O}_1\rangle_{J/\psi}$ was taken from
Ref.~\cite{Braaten:2002fi}, in which the relativistic correction to 
$\Gamma[J/\psi\to e^+e^-]$ was not taken into account. The inclusion of 
that correction in the present work, along with a more
precise determination of the potential-model parameter $\lambda$, 
accounts for the difference in the value of $\langle\bm{q}^2\rangle_{J/\psi}$
between Ref.~\cite{Bodwin:2006dn} and Eq.~(\ref{psi-me-nums}).

We can also compare our results with those in Ref.~\cite{He:2007te}. In 
that work, the following values are reported:
$\langle \mathcal{O}_1\rangle_{J/\psi}^{\textrm{HFC}} =0.573\hbox{~GeV}^3$,
$\langle \mathcal{O}_1\rangle_{\eta_c}^{\textrm{HFC}} =0.432\hbox{~GeV}^3$,
and $\langle \mathcal{P}_1\rangle_{J/\psi}^{\textrm{HFC}}/m_c^2
=\langle \mathcal{P}_1\rangle_{\eta_c}^{\textrm{HFC}}/m_c^2
=0.0514\hbox{~GeV}^3$. These values were obtained by 
comparing the theoretical formulas for 
$\Gamma[J/\psi\to e^+e^-]$,
$\Gamma[\eta_c\to\gamma\gamma]$, and
$\Gamma[J/\psi\to \textrm{light hadrons}]$ with the experimental 
results and by assuming that $\langle 
\mathcal{P}_1\rangle_{\eta_c}^{\textrm{HFC}}
=\langle \mathcal{P}_1\rangle_{J/\psi}^{\textrm{HFC}}$. 
In the case of $\Gamma[J/\psi\to \textrm{light hadrons}]$, 
processes involving an intermediate virtual photon were excluded in both 
the theoretical formula and the experimental rate. The theoretical 
expressions that were used in Ref.~\cite{He:2007te} to obtain these 
results contain the QCD corrections of relative order $\alpha_s$ and the
relativistic corrections of order $v^2$. Taking $m_c=1.5$~GeV, which is 
the value that is used in Ref.~\cite{He:2007te}, we find that the 
results given in Ref.~\cite{He:2007te} yield
$\langle\bm{q}^2\rangle_{J/\psi}^{\textrm{HFC}}=0.202$~GeV$^2$ and
$\langle\bm{q}^2\rangle_{\eta_c}^{\textrm{HFC}}=0.268$~GeV$^2$.
These values are considerably below the values in 
Eqs.~(\ref{psi-me-nums}) and (\ref{etac-gam-gam-me-nums})
and considerably below the expectations
from the velocity-scaling rules of NRQCD. The small values of 
$\langle\bm{q}^2\rangle_{J/\psi}^{\textrm{HFC}}$ and 
$\langle\bm{q}^2\rangle_{\eta_c}^{\textrm{HFC}}$ are traceable to the use 
of the theoretical expression for $\Gamma[J/\psi\to \textrm{light 
hadrons}]$. In that expression, the coefficient of the contribution that 
is proportional to $\langle\bm{q}^2\rangle_{J/\psi}/m_c^2$ is about $-5.32$ 
relative to the leading contribution. Because of this large negative 
coefficient, the quantity $\langle\bm{q}^2\rangle_{J/\psi}$ must be much 
less than the values that we obtain in order for the decay width to be 
positive. We regard this as an indication that the $v$ expansion is not 
reliable for the rate $\Gamma[J/\psi\to \textrm{light
hadrons}]$. It is possible that the resummation 
methods that we have used in the present work could be used to tame 
the $v$-expansion for $\Gamma[J/\psi\to \textrm{light hadrons}]$. The
value of $\langle \mathcal{O}_1\rangle_{J/\psi}^{\textrm{HFC}}$ is about
$30\%$ larger than the value in Eq.~(\ref{psi-me-nums}) while the value
of $\langle \mathcal{O}_1\rangle_{\eta_c}^{\textrm{HFC}}$ is about
$1\%$ smaller than the value in Eq.~(\ref{etac-gam-gam-me-nums}). Some
of this difference is accounted for by the smaller values of
$\langle\bm{q}^2\rangle_{J/\psi}$ and $\langle\bm{q}^2\rangle_{\eta_c}$
in Ref.~\cite{He:2007te}. Ref.~\cite{He:2007te} also makes use of
slightly different values of $m_c$ ($1.5$~GeV) and $\alpha_s$ ($0.26$)
than those employed in the present work. A further difference is
that the expressions for $\Gamma[J/\psi\to e^+e^-]$ and
$\Gamma[\eta_c\to\gamma\gamma]$ in Ref.~\cite{He:2007te} are expanded to
first order in $\alpha_s$ and $v^2$, rather than expressed as exact
squares of amplitudes, as in Eqs.~(\ref{G3H-F}) and (\ref{G1H-F}).

Finally, there are quenched lattice computations \cite{bks} of the
ground-state $S$-wave charmonium matrix elements that yield $\langle
\mathcal{O}_1\rangle_{J/\psi\hbox{-}\eta_c}=0.3312\pm 0.0006\pm
0.0030_{-0.0483}^{+0.0681}$~GeV$^3$ and
$\langle\bm{q}^2\rangle_{J/\psi\hbox{-}\eta_c}=%
0.07\,\hbox{--}\,0.82$~GeV$^2$. In $\langle
\mathcal{O}_1\rangle_{J/\psi\hbox{-}\eta_c}$, the first error bar is
from lattice statistics, the second error bar is from lattice
systematics, and the third error bar is from the uncertainty in the
one-loop perturbative computation that relates the lattice-regulated
matrix elements to the continuum $\overline{\textrm{MS}}$ matrix
elements. The lattice computations do not distinguish between the
$J/\psi$ state and the $\eta_c$ state. The lattice results
are in agreement with our results, within uncertainties, but the lattice
uncertainties are much larger than ours. These large uncertainties
arise from the uncertainty in the perturbative conversion from lattice
to continuum $\overline{\textrm{MS}}$ matrix elements.

\section{Summary\label{sec:conclusions}}

For many $S$-wave heavy-quarkonium decay and production processes, 
the color-singlet $S$-wave NRQCD matrix elements of 
leading order in $v$ enter into the dominant theoretical contribution. 
The first relativistic corrections to these processes involve the matrix 
elements of relative order $v^2$. 

We have computed the color-singlet $S$-wave NRQCD matrix elements of
leading order and next-to-leading order in $v^2$ for the $J/\psi$ and
the $\eta_c$. For each of these quarkonium states, we have determined
the values of these matrix elements by comparing the theoretical
expressions for the electromagnetic decay rates ($\Gamma[J/\psi\to
e^+e^-]$ or $\Gamma[\eta_c\to \gamma\gamma]$) with the experimental
measurements and by using a potential model to compute the matrix
elements of relative order $v^2$. If the static, spin-independent $Q
\bar Q$ potential were known exactly, then the potential-model
calculation would be accurate up to corrections of relative order $v^2$.
We made use of the Cornell potential and fixed its parameters by using as
inputs the lattice measurements of the string tension, the
$J/\psi$-$\psi(2S)$ mass splitting, and the quarkonium wave function at
the origin, which corresponds to the NRQCD matrix element of leading
order in $v$. Because the potential-model calculation of the order-$v^2$
matrix element depends on the leading-order NRQCD matrix element and the
decay widths depend on both of these matrix elements, we obtained the
matrix elements for the $J/\psi$ and the $\eta_c$ by solving, in each
case, two coupled nonlinear equations.

In the theoretical expressions for the electromagnetic decay widths, we
made use of the generalized Gremm-Kapustin relation
(\ref{gremm-kapustin}) (Ref.~\cite{Bodwin:2006dn}) to resum a class
of relativistic corrections. This resummation includes
all of the relativistic corrections that are contained in the 
leading-potential approximation to the quarkonium $Q\bar Q$ color-singlet 
wave function, up to the ultraviolet cutoff of the NRQCD matrix 
elements. 

There are many sources of uncertainties in our calculation. Some of
these are correlated among the matrix elements. Therefore, we
reported the variations of the matrix elements with respect to each
source of uncertainty.

The experimental measurement of the width $\Gamma[\eta_c\to
\gamma\gamma]$ has relatively large uncertainties, which translate into
large uncertainties in the $\eta_c$ matrix elements. Owing to the
heavy-quark spin symmetry \cite{Bodwin:1994jh}, the $J/\psi$ and
$\eta_c$ matrix elements are equal, up to corrections of relative order
$v^2$. Therefore, we were able to reduce the uncertainties in the
$\eta_c$ matrix elements by averaging the values that we obtained from
$\Gamma[\eta_c\to \gamma\gamma]$ plus the potential model with the
values that we obtained from $\Gamma[J/\psi\to e^+e^-]$ plus the
potential model. In performing this average, we took into account the
additional uncertainty of relative order $v^2$ that arises from
equating $\eta_c$ matrix elements to $J/\psi$ matrix elements.

Our principal results are given in Tables~\ref{table1}, \ref{table2},
and \ref{table3} and are summarized in Eqs.~(\ref{psi-me-nums}),
(\ref{etac-gam-gam-me-nums}), and (\ref{etac-me-with-mc}) for the
matrix elements that were determined from $\Gamma[J/\psi\to e^+e^-]$,
the matrix elements that were determined from $\Gamma[\eta_c\to
\gamma\gamma]$, and the average of the two, respectively. We consider
the results in Table~\ref{table3} and Eq.~(\ref{etac-me-with-mc}) to be
our best values for the $\eta_c$ matrix elements. In applying these
results to calculations of quarkonium decay and production rates, it
should be kept in mind that the uncertainties are highly correlated
between matrix elements and that there are correlations between matrix
elements and short-distance coefficients with respect to the uncertainties
in $m_c$. Therefore, it may be necessary to use all of the information
that is contained in Tables~\ref{table1}, \ref{table2}, and
\ref{table3}, rather than to rely on the summaries in
Eqs.~(\ref{psi-me-nums}), (\ref{etac-gam-gam-me-nums}), and
(\ref{etac-me-with-mc}).

Our results in Tables~\ref{table1} and \ref{table2} and
Eqs.~(\ref{psi-me-nums}) and (\ref{etac-gam-gam-me-nums}) conform to the
expectation, from the heavy-quark spin symmetry, that the $J/\psi$ and
$\eta_c$ matrix elements are equal, up to corrections of relative order
$v^2\approx 30\%$. In fact, the leading-order matrix elements differ by
about $1.5\%$, while $\langle \bm{q}^2\rangle_{J/\psi}$ and $\langle
\bm{q}^2\rangle_{\eta_c}^{\gamma\gamma}$ differ by only about $0.5\%$.
The velocity-scaling rules of NRQCD \cite{Bodwin:1994jh} state that the
quantities $\langle v^2\rangle_H =\langle\bm{q}^2\rangle_H/m_c^2$
should be of order $v^2\approx 0.3$. From Eq.~(\ref{vsq}), it can be
seen that our results satisfy this expectation, although they are
somewhat smaller than the nominal value of $v^2$.

As we have discussed in Sec.~\ref{basic-comps}, the effects from
resummation on our results are small, ranging from 
$-5.7\%$ for $\langle\mathcal{O}_1\rangle_{\eta_c}^{\gamma\gamma}$ 
to $2.1\%$ for $\langle\bm{q}^2\rangle_{\eta_c}^{\gamma\gamma}$.
The small effects from
resummation suggest that the $v$ expansion of NRQCD converges well for
the widths $\Gamma[J/\psi\to e^+e^-]$ and $\Gamma[\eta_c\to\gamma\gamma]$.

Our results for $\langle\mathcal{O}_1\rangle_{J/\psi}$ and 
$\langle\mathcal{O}_1\rangle_{\eta_c}^{\gamma\gamma}$ are considerably
larger than those in Ref.~\cite{Braaten:2002fi}, primarily because we 
have included relativistic corrections to the electromagnetic decay 
rates in the present work. The changes in the values of these matrix 
elements would significantly increase the rate for the process 
$e^+e^-\to J/\psi+\eta_c$ that is calculated in Ref.~\cite{Braaten:2002fi}.

Our result for $\langle\bm{q}^2\rangle_{J/\psi}$ agrees, within
uncertainties, with that in Ref.~\cite{Bodwin:2006dn}, but is slightly
smaller. Most of this difference arises from the fact that, in
Ref.~\cite{Bodwin:2006dn}, the value of
$\langle\mathcal{O}_1\rangle_{J/\psi}$ was taken from
Ref.~\cite{Braaten:2002fi}.

In Ref.~\cite{He:2007te}, a much smaller value for
$\langle\bm{q}^2\rangle_{J/\psi}$ was reported.
($\langle\bm{q}^2\rangle_{\eta_c}$ was assumed to be equal to
$\langle\bm{q}^2\rangle_{J/\psi}$ in this work.) The smallness of
$\langle\bm{q}^2\rangle_{J/\psi}$ in Ref.~\cite{He:2007te} can be
traced to the use of the width
$\Gamma[J/\psi\to \textrm{light hadrons}]$ to constrain the matrix
elements. The theoretical expression for that width contains large 
order-$v^2$ corrections that, in our opinion, make the reliability of the
expression suspect. It is possible that the resummation technique that
we have employed in this paper could be used to bring the $v$ expansion
for $\Gamma[J/\psi\to \textrm{light hadrons}]$ under control.

Our results for $\langle\mathcal{O}_1\rangle_{J/\psi}$ and 
$\langle\bm{q}^2\rangle_{J/\psi}$ are in agreement with those from 
lattice calculations \cite{bks}, although the lattice uncertainties are 
much larger than ours.

We believe that the values that we have obtained for the $J/\psi$ and
$\eta_c$ color-singlet NRQCD matrix elements are the most precise ones
that are available to date. The new values for the matrix elements of
leading order in $v$ should have a significant impact on the
calculations of a number of charmonium decay and production
processes~\cite{BLY}. For quite a few charmonium processes, it is clear
that relativistic corrections are important. Within the framework of
NRQCD, the matrix elements of order $v^2$ are essential ingredients in
calculating those corrections. We have also attempted to quantify all of
the significant theoretical uncertainties in our determination of the
$J/\psi$ and $\eta_c$ color-singlet matrix elements. Our treatment of
uncertainties could provide the basis for more reliable estimates of
theoretical uncertainties in future calculations of charmonium decay and
production rates.

\begin{acknowledgments}
We thank Jens Erler for providing us with the latest version of the 
code GAPP and for explaining its use. 
JL thanks the High Energy Physics Theory Group at Argonne National
Laboratory for its hospitality while this work was carried out.
Work by GTB in the High Energy Physics Division at Argonne
National Laboratory is supported by the U.~S.~Department of Energy,
Division of High Energy Physics, under Contract No.~DE-AC02-06CH11357.
The work of HSC was supported by the Korea Research Foundation (KRF) 
under MOEHRD Basic Research Promotion grant KRF-2006-311-C00020.
The work of DK was supported by the Korea Science and Engineering 
Foundation (KOSEF) under grant R01-2005-000-10089-0.
The work of JL was supported by KRF under grant KRF-2004-015-C00092
and by a Korea University Grant.
The work of CY was supported by 
KRF funded by Korea Government (MOEHRD, Basic Research Promotion Fund) 
(KRF-2005-075-C00008).
\end{acknowledgments}

{}


\end{document}